\documentclass[%
 reprint,
superscriptaddress,
 amsmath,amssymb,
 aps,
]{revtex4-1}

\usepackage{graphicx}
\usepackage{dcolumn}
\usepackage{bm}
\usepackage{braket}
\usepackage{natbib}

\begin{document}

\preprint{APS/123-QED}

\title{Mixed State Dynamics with Non-Local Interactions}

\author{Erik W. Lentz}
\author{Leon Lettermann}
\affiliation{Institut f\"ur Astrophysik, Georg-August Universit\"at G\"ottingen, 
                 G\"ottingen, Deutschland 37077;       
		  {\tt erik.lentz@uni-goettingen.de,
		  leonlettermann@gmail.com}}
\author{Thomas R. Quinn}
\affiliation{Astronomy Department, University of Washington, 
                 Seattle, WA 98195-1580;       
		  {\tt trq@astro.washington.edu}}
\author{Leslie J Rosenberg}
\affiliation{Physics Department, University of Washington,
                 Seattle, WA 98195-1580;       
		 {\tt lentze@phys.washington.edu, ljrosenberg@phys.washington.edu}}

\date{\today}

\begin{abstract}
The evolution of degenerate matter out of equilibrium is a topic of interest in fields such as condensed matter, nuclear and atomic physics, and increasingly cosmology, including inflaton physics prior to reheating. This follow-up paper extends the recent paper on the super-de Broglie dynamics of pure condensates of non-relativistic identical particles subject to non-local two-body interactions to the dynamics of mixed states. It is found that the two-body correlation function plays an increasingly dynamical role in these systems, driving the development of condensates and distributed phases alike. Examples of distribution and correlation evolution are presented, including instances of collapse, bound and unbound states, and phonons in the bulk. Potential applications are also discussed.

\end{abstract}

\pacs{Valid PACS appear here}
\maketitle


\section{\label{sec:Introduction}Introduction}

The recent paper by \citet{Lentz2018a} (LQR) uncovered the role of particle exchange and inter-particle correlation in the dynamics of condensed fluids with non-local interactions on super-de Broglie lengths. The approach covered a wide range of interactions, including those with infinite-range such as the Coulombic potential of electrostatics and Newtonian gravity. The derived model of dynamics found that such Bose and Fermi systems exhibited extra-classical behavior in the form of exchange-correlation interactions, even above the de Broglie scale. The resulting equation of motion for the condensate is concise and highly tractable, making it an attractive model. For instance, the authors of LQR have used this model to describe cosmological structure formation of an axion dark matter candidate \cite{Lentz2018b,Lentz2018c}.

A pure condensed state is often too constraining of a description for many degenerate systems, however, particularly for states out of equilibrium. The process of forming or destroying a condensate and the interaction of multiple phases of a degenerate fluid are examples of mixed systems that are beyond the scope of LQR's derivation. This paper updates the previous derivation to allow for the occupation of a density of states, greatly enlarging the domain of utility. Exchange-correlation (XC) interactions are again seen to play a central and increasingly dynamical role in the presentation of interaction forces on the fluid's state. The resulting equations of motion for the fluid are found. Relations to the condensed and diffuse limits are then discussed. Intuitive and tangible examples of this model are also presented to demonstrate features of Bose and Fermi fluids in the presence of attractive, repulsive, and mixed interactions.

\section{\label{sec:Derivation}Derivation}

The many-body quantum mechanical system presented in LQR consists of $N$ identical Fermi or Bose particles, restricted to states that are symmetric over all non-spatial degrees such as spin. Further, the solution space is to be restricted to states that can be expressed as a product of spatial and non-spatial degrees, where the non-spatial degrees of freedom are in symmetric form. The Schr\"odinger equation expressed in the spatial basis then takes the form
\begin{equation}
i \hbar \partial_t  \Psi \left( \vec{x}_1,...,\vec{x}_N;t \right) = H \Psi \left( \vec{x}_1,...,\vec{x}_N;t \right),
\end{equation}
where the Hamiltonian operator is represented by
\begin{equation}
H = - \sum_i^N \frac{\hbar^2 \nabla^2_i}{2 m} + \sum_{i < j}^N \phi \left(|\vec{x}_i-\vec{x}_j| \right),
\end{equation}
where the first term is the canonical non-relativistic kinetic-energy contribution from each particle and the second term represents the non-local inter-particle interactions, parameterized by the single smooth, real potential function $\phi$. Only non-local central interactions are considered here. The system of identical particles also exhibits (anti-)symmetry under particle exchange, written as 
\begin{align}
\Psi \left( \vec{x}_1,...,\vec{x}_i,...,\vec{x}_j,...,\vec{x}_N;t \right) &= \mp \Psi \left( \vec{x}_1,...,\vec{x}_j,...,\vec{x}_i,...,\vec{x}_N;t \right) \nonumber \\ 
& \forall i,j
\end{align}
for fermions and bosons, respectively.

The mixed state of the many-body system can be tracked using the density matrix operator
\begin{equation}
\hat{\rho} = \sum_{\alpha \in A }p_{\alpha}(t) \ket{\alpha} \bra{\alpha},
\end{equation}
where $p_{\alpha}(t)$ is the probability of occupying in the $\ket{\alpha}$ state, and $A$ is the non-degenerate set of orthonormal vectors of the N-body Hilbert space. The evolution of the density operator is found via the operator form of Liouville's theorem
\begin{equation}
\dot{\hat{\rho}} = \frac{1}{i \hbar} \left[\hat{\rho},\hat{H}\right].
\end{equation}
The expected evolution of an operator $\hat{O}$ over the ensemble is then
\begin{equation}
\partial_t \braket{\hat{O}} = \frac{1}{i \hbar} \mathrm{Tr} \left( \hat{\rho} \left[\hat{O},\hat{H}\right]\right). \label{eqn:opevol}
\end{equation}

A many-body Wigner transform operator is constructed to describe the (quasi-)distribution of the system's states over the $N$ phase spaces $w_i = (\vec{x}_i,\vec{p}_i)$
\begin{align}
&\hat{f}^{(N)}(w_1,...,w_N,t) = \int d^3x'_1 \cdot \cdot \cdot d^3x'_N e^{i \sum_j^N \vec{p}_j \cdot \vec{x}'_j/\hbar} \times \nonumber \\
& \hat{\Psi}^{\dagger}(\vec{x}_1+\vec{x}'_1/2,...,\vec{x}_N+\vec{x}'_N/2,t)\times \nonumber \\
&\hat{\Psi}(\vec{x}_1-\vec{x}'_1/2,...,\vec{x}_N-\vec{x}'_N/2,t),
\end{align}
which will be used to track the motions of the fluid. The evolution of the Wigner operator's ensemble expectation $f^{(N)} = \braket{ \hat{f}^{(N)}}$ can be calculated using Eqn.~\ref{eqn:opevol} and is found to obey the many-body Liouville equation to lowest order in $\hbar/m$,
\begin{equation}
\partial_t f^{(N)} + \sum_i^N \frac{\vec{p}_i}{m} \vec{\nabla}_i f^{(N)} - \sum_{i \ne j}^N \vec{\nabla}_i \phi_{ij} \cdot \vec{\nabla}_{p_i} f^{(N)} = \mathcal{O}(\hbar/m), \label{eqn:DFEOM}
\end{equation}
where the use of the zeroth order terms puts us firmly in the super-de Broglie limit, also consistent with the Husimi distribution with an envelope such that $\sigma_x^2 \sigma_p^2$ is larger than $\hbar^2$ \citep{Husimi1940}. An additional advantage of operating in this limit is that the expected many-body Wigner function can be interpreted as a true distribution function (DF).

The simplification of the mixed N-body function is not as clear as for the condensate presented in LQR, as the Runge-Gross theorem \cite{Runge1984} no longer applies. We may still integrate out phase spaces $2,...,N$ of $f^{(N)}$ to find the motion of the single-body density
\begin{equation}
0=\partial_t f^{(1)} + \frac{\vec{p}_1}{m} \cdot \vec{\nabla}_1 f^{(1)}  - (N-1)\int d^6w_2 \vec{\nabla} \phi_{12} \cdot \vec{\nabla}_{p_1} f^{(2)},
\end{equation}
though the two-body DF is no longer guaranteed to be functionally described in terms of the single-body DF. 

The presence of correlations between bodies found in the many-body wave functions presented in LQR direct us to consider non-trivial correlations in the mixed system as well, though the behavior of the correlations is less clear. Determining the two-body DF equation of motion is done here by integrating out phase-spaces $3,...,N$ from Eqn.~\ref{eqn:DFEOM}
\begin{align}
0=\partial_t f^{(2)} +& \sum_{i=1}^2\Bigg[\frac{\vec{p}_i}{m} \cdot \vec{\nabla}_i f^{(2)} - \vec{\nabla}_i \phi_{12} \cdot \vec{\nabla}_{p_i} f^{(2)} \nonumber \\
&- (N-2)\int d^6w_3 \vec{\nabla}_i \phi_{i3} \cdot \vec{\nabla}_{p_i} f^{(3)}\Bigg],
\end{align}
which depends on the expected three-body DF $f^{(3)}$. 
In this system it is possible to truncate this hierarchy of dependency by noting the exchange relations and structure of the two-body pairwise interactions permit the reduction of the three-body DF via the Kirkwood expression \citep{Pearl1988}
\begin{align}
    & f^{(3)}(w_1,w_2,w_3,t) = \nonumber \\
    &\frac{f^{(2)}(w_1,w_2,t) f^{(2)}(w_2,w_3,t) f^{(2)}(w_3,w_1,t)}{f^{(1)}(w_1,t) f^{(1)}(w_2,t) f^{(1)}(w_3,t)}.
\end{align}
One can also write the Kirkwood expression in terms of the dimensionless two-body correlation function $\tilde{g}(w_1,w_2,t) = f^{(2)}(w_1,w_2,t)/ f^{(1)}(w_1,t) f^{(1)}(w_2,t)$. The super-de Broglie dynamics of the mixed many-body system can therefore be reduced to the solutions of the single-body DF and the two-body correlation function
\begin{align}
&0=\partial_t f^{(1)}_1 + \frac{\vec{p}_1}{m} \cdot \vec{\nabla} f^{(1)}  \nonumber \\
&- \frac{N-1}{N}\int d^6w_2 \vec{\nabla}_1 \Phi_{12} \cdot \vec{\nabla}_{p_1} \left(f^{(1)}_1 \tilde{g}_{12} f^{(1)}_2 \right) , \label{DF1EOM} \\
&0=\partial_t \tilde{g}_{12}  + \sum_{i=1}^2\Bigg[ \frac{\vec{p}_i}{m} \cdot \vec{\nabla}_i \tilde{g}_{12} \nonumber \\
&- \frac{N-2}{N f^{(1)}_i}\int d^6w_3 \vec{\nabla}_i \Phi_{i3} \cdot  \Bigg( \vec{\nabla}_{p_i} \left( \tilde g_{12}\tilde g_{13}\tilde g_{23} f^{(1)}_i f^{(1)}_3  \right) \nonumber \\
&- \tilde g_{12} \vec{\nabla}_{p_i}\left(\tilde{g}_{i3} f^{(1)}_i f^{(1)}_3 \right) \Bigg)\Bigg],  \label{CorrEOM}
\end{align}
where $\Phi_{12} = N \phi_{12}$, and subscripts on correlation functions and single-body DFs indicate the phase spaces of their arguments. The trivial correlation limit ($\tilde g=1$) is seen to be a stationary point of the solution space, in the absence of external potentials. The condensed limit shown in LQR is also seen to be a potential solution of the generalized system, though the stability of any one such solution depends on the type of interaction and the correlation of other available states, condensed and not.

\section{\label{sec:Demonstrations} Examples}

This section presents several examples of the model described in Eqns.~\ref{DF1EOM}, \ref{CorrEOM}. Systems are chosen so that they are solvable either analytically or by using standard numerical ODE packages. Examples will also operate within the limits of small correlation perturbations $|\delta g| = |1- \tilde{g}| \ll 1$, and with sufficiently many particles that $(N-1)/N \to 1$.

\subsection{\label{sec:shells} Dynamics of Spherical Shells}

The first examples of this method are extensions of the example given in LQR, covering the collapse of cold spherical shells under self-gravitation, expansion under electrostatic repulsion, and oscillation in a bound state of a macroscopic van der Waals-like force. For simplicity, we again assume no tangential fluid motion, that the radial velocity dispersion is sufficiently small so as to leave the shell width unchanged over the simulated collapse, and that the two-body correlation function can be approximated as homogeneous over the shell. The ansatz for the one-body distribution is $f^{(1)}_{shell}(r,m v_r,t) = A \times \mathrm{exp} \left[ - (r-r_c)^2/ 2 \sigma_r^2 - (v_r - \dot{r}_c)^2/ 2 \sigma_v^2   \right]$, where $r_c$ is the shell center radius, $A$ is the DF normalization, $\sigma_r$ is the thickness of the shell, and $\sigma_v$ is the velocity dispersion of the shell. These assumptions reduce the complexity of the problem to the degrees of the shell's radius and the deviation of the correlation function from unity, $r_c(t)$, $\delta g(t)$.

The governing equation of the shell under an interaction force profile $- \partial_r \Phi = F $ may then be written to leading order in dispersion and correlation perturbation as
\begin{align}
\ddot{r}_c &= F(r_{soft}) \left(1 + \delta g \right), \\
\dot{\delta g} &= - 2 F(r_{soft}) \frac{\delta g}{\sigma_v},
\end{align}
where $r_{soft}^2 = \left(r^2_c + \sigma_r^2\right)$. The factor of two in the correlation equation of motion relates to the two phase spaces by which interaction enters the correlation equation of motion.

\subsubsection{\label{sec:Demonstration1} Collapse of Self-Gravitating Shell}

\begin{figure*}[ht!]
\begin{center}
\includegraphics[width=\textwidth]{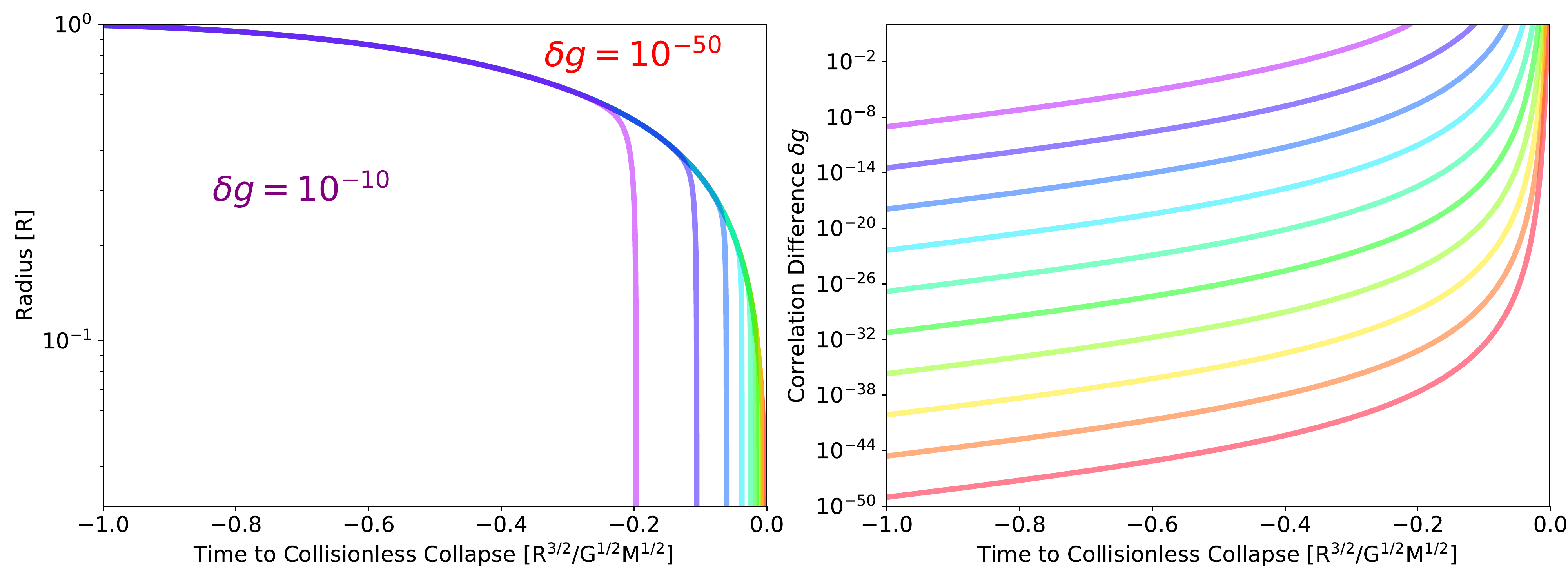}
\caption{Demonstration of dynamical correlation on collapse of spherical shell. (Left) Radial trajectories of collapsing self-gravitating spherical shells of Bose fluid. (Right) Correlation perturbation evolution during collapse. Shells start from rest. The color gradient is logarithmic in initial correlation perturbations $\delta g_0 \in [10^{-50}, 10^{-10}]$, which cover a large range as the expected size of cosmological correlations at the start of collapse are as of yet unknown. The XC physics alters the infall of the shell, generally increasing the attraction and making the collapse more violent. The collapse also increases the size of the correlation, creating positive feedback. Near to collapse, the correlation exits the $\delta g \ll 1$ regime, compromising the reliability of the results at collapse.}
\label{sph_shell_coll}
\end{center}
\end{figure*}

The collapsing spherical shell is subject to the Newtonian gravitational force
\begin{equation}
    F(r) = -\frac{GM}{r^2},
\end{equation}
where $G$ is the Newton gravitational constant, and $M$ is the self-gravitating mass of the shell. A relevant example would be a cold dark matter candidate of a sufficiently small particle mass, such as the axion, that is expected to be extraordinarily degenerate at the presumed dark matter density and temperature \cite{Marsh2016,Guth2015,Sikivie2009}. Solutions to the collapse of an initially stationary sphere for a variety of perturbed correlations show that the rate of infall is amplified by correlation and that attractive infall amplifies the degree of correlation in a cycle of positive feedback, see Fig. \ref{sph_shell_coll}. The case of zero initial correlation (classical mean field theory) remains without correlation throughout, but the mean field is an unstable stationary point among attractive interactions. One can also see a similar behavior among fermions ($\delta g \le 0$), with collapse causing growing anti-correlations that slow the rate of collapse. Parameters chosen for Fig. \ref{sph_shell_coll} are $r(t=0)=1$, $\sigma_r=10^{-3}$, and $\sigma_v=10^{-1}$ in dynamical units. The growth of correlation under self-gravitation seems to contradict to the findings of \citet{Guth2015} that long-distance correlations are quenched by collapse.

\subsubsection{\label{sec:Demonstration2} Quenching of Repulsed Shells}

\begin{figure*}[ht!]
\begin{center}
\includegraphics[width=\textwidth]{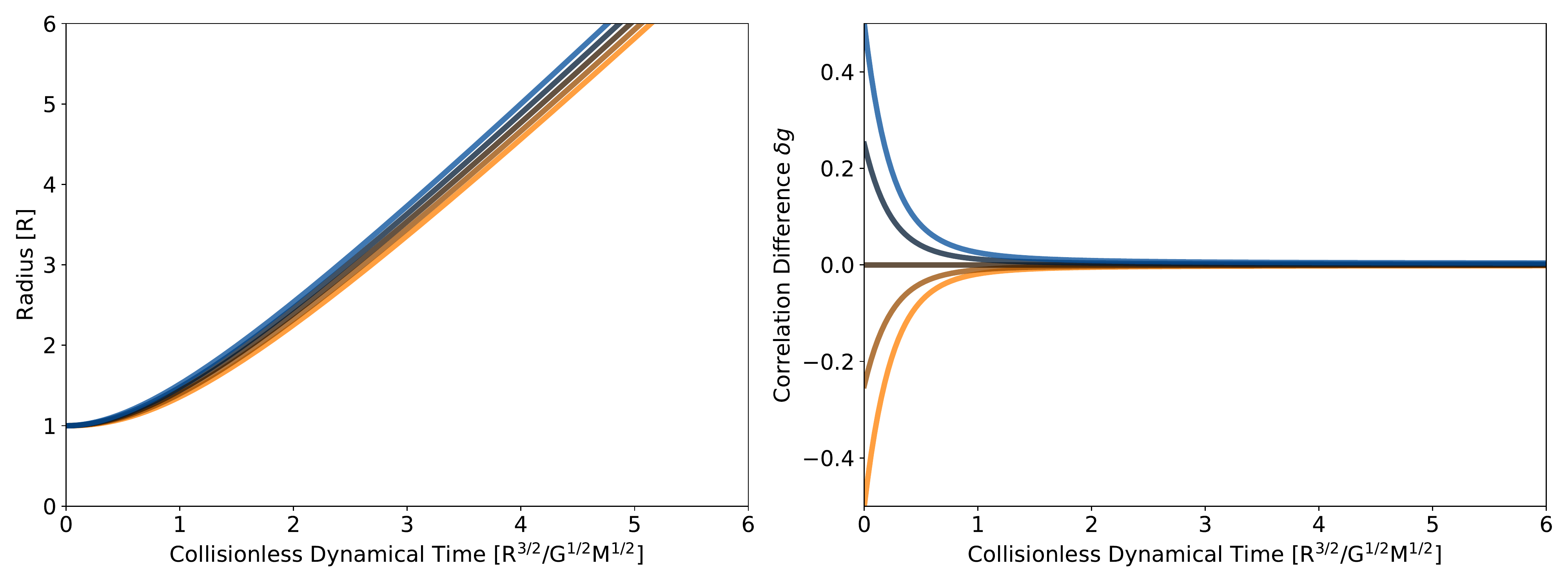}
\caption{Demonstration of dynamical correlation on expansion of spherical shell under Coulombic repulsion. (Left) Radial trajectories of spherical shells of Bose and Fermi fluids. (Right) Correlation perturbation evolution during expansion. Shells start from rest. The color gradient is linear in initial correlation perturbations $\delta g_0 \in [-0.5,0.5]$, with Bose fluids having non-negative correlation perturbations and Fermi fluids having non-positive correlation perturbations. The XC physics alters the magnitude of the Coulombic force, altering the exit velocity of the shell. The repulsive force quenches the correlation perturbation for both fluid types.}
\label{sph_shell_repul}
\end{center}
\end{figure*}

If the same spherical shell is subject to a long-range repulsive interaction, such as the Coulomb electrostatic force
\begin{equation}
    F(r) = \frac{k C^2}{M r^2},
\end{equation}
where $k$ is the Coulomb constant and $C$ is the effective charge, the shell will not collapse but instead expand, with solutions shown in Fig.~\ref{sph_shell_repul}. The relevant example here would be a shell of cold ionic gas released from a trap. Correlations are seen to change the time of flight of the shell towards a particular radius. The dimensionless correlation perturbations also change their behavior, decaying towards zero for both Bose and Fermi systems. The classical  mean field theory limit of $\tilde{g} = 1$ is now a stable stationary point. Such an example fulfills the intuition that (repulsive) interactions quench correlations.

\subsubsection{\label{sec:Demonstration3} Stable Bound States}

\begin{figure*}[ht!]
\begin{center}
\includegraphics[width=\textwidth]{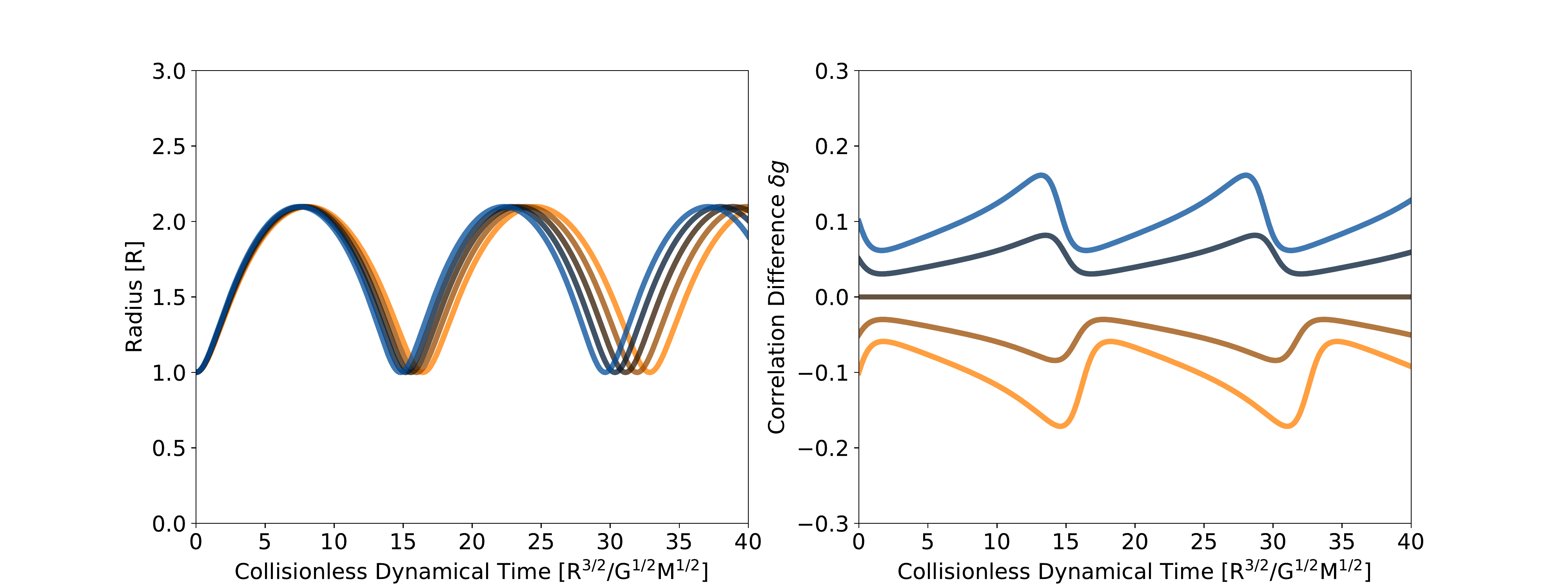}
\caption{Demonstration of dynamical correlation on the motion of a bound spherical shell subject to a Van der Waals force. (Left) Radial trajectories of spherical shells of Bose and Fermi fluids. (Right) Correlation perturbation evolution during oscillation. Shells start from rest. The color gradient is linear in initial correlation perturbations $\delta g_0 \in [-0.1,0.1]$, with Bose fluids having non-negative correlation perturbations and Fermi fluids having non-positive correlation perturbations. }
\label{sph_shell_VdW}
\end{center}
\end{figure*}

The final shell example is for a van der Waals-like force
\begin{equation}
    F(r) = \frac{G M}{R} \left( \frac{-1}{r^3} + \frac{1.3 R}{r^4} \right),
\end{equation}
where $r=R$ is the initial radius of the shell. The initial condition of $r_c(t=0)=1$ places the shell in a classical bound state sufficiently far from the ground state at $r=1.3 R$. A relevant example may be a shell of neutral cold gas. The resultant oscillatory bound state is stable for small correlations, with correlations taking their extremal values as the shell passes through the interaction potential minimum, while at other times growing or shrinking in amplitude as the shell experiences an attractive or repulsive force respectively, as shown in Fig.~\ref{sph_shell_VdW}. Interestingly, the amplitude of the shell oscillation appears unchanged by the size of the seed correlation. Seed correlation appears to only change the period of the shell's oscillation.

\subsection{\label{sec:bulk} Bulk Speed of Sound}

The impact of correlations on the bulk properties of Bose and Fermi fluids is also interesting. Consider perturbative sound waves propagating through a large bulk system subject to a finite-range repulsive force. For wavelengths much longer than the inter-particle spacing and the scattering length of the interaction, the effective potential becomes point-like, behaving in the kinetic regime as $\sim a \rho$, where $a$ is the interaction scale. The equations of motion for the density perturbations and correlation perturbations are then found to be
\begin{align}
\ddot{\delta}(\vec{x}_1,t) &= c_s^2 \nabla_1^2 \delta(\vec{x}_1,t) + \frac{c_s^2}{\rho} \nabla_1^2 \mu (\vec{x}_1,\vec{x}_1,t), \\
\ddot{\mu} (\vec{x}_1,\vec{x}_2,t) &= c_s^2 \nabla_1^2 \mu (\vec{x}_1,\vec{x}_2,t) + c_s^2 \nabla^2_2 \mu (\vec{x}_1,\vec{x}_2,t),
\end{align}
where $\delta$ is the spatial density perturbation, $\mu$ is a form of the spatial two-body density perturbation with units of density squared, $\rho$ is the mean bulk density, and $c_s$ is the base sound speed of the correlation-less fluid. Note that perturbations in the correlation function can propagate in two directions. The density perturbations can be considered as a propagating mode in the limit of small $\mu/\rho \delta$. The speed of sound of density perturbations is found to be 
\begin{equation}
    c'_s = c_s \sqrt{1+ \frac{\mu_0}{\delta_0 \rho}},
\end{equation}
where $\delta_0$ is initial amplitude of the density perturbations and $\mu_0$ gives the homogeneous correlation displacement and the wave's initial intitial correlation. The presence of correlation is seen to increase the speed of sound in Bose fluids relative to the mean field and decrease the speed of sound in Fermi fluids.

\section{\label{sec:Conclusions}Conclusions}

This paper presents a concise model of macroscopic dynamics for mixed systems of identical particles with non-local two-body interactions, expanding on the condensate dynamics presented in LQR. A density matrix approach was used to find the evolution of a mixed system of $N$ non-relativistic bodies. Those dynamics were transformed into a Wigner function, and its equation of motion was derived in the super-de Broglie limit. A closed system of Boltzmann-like field equations involving the single-body distribution function and the two-body correlation function was found to govern the many-body system. Deviations from the classical and standard mean-field descriptions were again found to be sourced by inter-body correlations, which now have their own dynamics. Correlations between particles are consistent with expected behavior in the condensed and diffuse limits. 

Examples of the model were made using a thin cold spherical shell and sound waves on a homogeneous bulk, both operating with small correlations. Attractive forces such as Newtonian self-gravity were seen to grow correlations over the course of collapse, while repulsive forces such as the electrostatic Coulombic force quenched correlations as the shell expanded. A shell caught in the well of a macroscopic van der Waals-like interaction observed both of these behaviors as the shell oscillated about the potential minimum, growing and shrinking its correlations as it is subjected to attractive and repulsive forces, respectively; the net effect being an alteration in the period of oscillation. Growth or decay of correlations are seen to be quite rapid, owing to the multiple channels available to the two-body interactions in the correlation's governing equation. The speed of sound in bulk was seen to be impacted by the presence of correlations. Sound speed was seen to be decreased by XC for Fermi fluids and increased for Bose fluids. All four examples contained a stationary point at the mean field value of correlation $\tilde{g} = 1$, though the repulsive/attractive nature of the interaction force is capable of changing the stationary point from stable to unstable. The seeding of correlations does not appear possible with the stated model, though external interactions are expected to be the remedy to this. The authors' current application of the mixed degenerate system to Bose (axion-like) dark matter shows great promise for the genesis and growth of correlations that are first induced by gravitational interactions with the photon fluid and then build by self-gravitation during the radiation era of cosmology. There are many other potential applications of this technique. For example, researchers in condensed matter, nuclear astrophysics, cold atomic physics, and others may find the above description useful. 

\section{\label{sec:Acknowledgements}Acknowledgements}

We would like to thank Bodo Schwabe, Sebastian Hoof, Mona Dentler, Viraf Mehta, and Fabian Heidrich-Meisner for their productive discussions in the refinement of this paper. We also gratefully acknowledge the support of the U.S. Department of Energy Office of High Energy Physics and the National Science Foundation. TQ was supported in part by the NSF grant AST-1514868. LR was supported in part by the DOE grant DE-SC0011665.


\bibliographystyle{apsrev4-1}
\bibliography{Bibliography.bib}

\end{document}